\documentclass[twocolumn,showpacs,preprintnumbers,amsmath,amssymb]{revtex4}
\usepackage{graphicx}% Include figure files
\usepackage{dcolumn}% Align table columns on decimal point
\usepackage{bm}% bold math

\begin {document}

\preprint{APS/123-QED}

\title{Magnetic-oscillation mechanism for understanding periodic or quasi-periodic modulation of the timing residuals from pulsars}

\author{Zhaojun Wang}
\email{wzj@xju.edu.cn}
 \altaffiliation[]{}%Lines break automatically or can be forced with \\
\author{Guoliang L\"{u}}%
 \email{guolianglv@xao.ac.cn}
 \author{Chunhua Zhu}%
 \author{Lin Li}
\affiliation{School of Physical Science and Technology,
Xinjiang University, Urumqi, 830046, China
}
 \author{Anzhong Wang}%
\affiliation{GCAP-CASPER, Physics Department, Baylor University, Waco, Texas 76798-7316, USA\\
Institute for Advanced Physics and Mathematics, Zhejiang University of Technology, Hangzhou, 310032, China}

%\affiliation{
\date{\today}% It is always \today, today,

\begin{abstract}
Highly precise pulsar timing is very important for understanding the nature of neutron stars and the implied physics, even being used to detect directly gravitational waves. Unfortunately, the accuracy of the pulsar timing is seriously affected by the spin-down irregularities, such as spin fluctuations with a manifestation
of low frequency structures (the so-called red noise processes), and various activities of magnetospheres. Except from the random timing noise, significant periodic
or quasi-periodic components could also be found in some timing residuals of pulsars, indicating the presence of unmodelled deterministic effects. The physical origins of these effects still remain unexplained. In this paper, we suggest a new mechanism involving the de Haas-van Alphen magnetic oscillation, which could trigger the observed low frequency structures. We find that the  proposed magnetic oscillation period is about 1-$10^{2}$ yr, which is about
$10^{-4}$ times as long as the classical characteristic time scale of interior magnetic field evolution for a normal neutron star. Due to the
magnetic oscillation, we estimate both the braking index range between $10^{-5}$ and $10^{5}$ and the residuals range between 4 to 906 ms for some  specific samples of quasi-periodic pulsars. Those are consistent with the pulsar timing observations. Nonlinear phenomena of de Haas-van Alphen effect and the Condon domain structure
are considered. Avalanche-like magnetized process of domain wall motion (rapidly magnetic energy release)
could associated with some special emitting manifestations of neuron star radiation.

%Valid PACS numbers may be entered using the \verb+\pacs{#1}+ command.
\end{abstract}

\maketitle

%\section{\label{sec:level1}First-level heading:\protect\\ The line
%break was forced \lowercase{via} \textbackslash\textbackslash}
\section{Introduction}
Neutron star (NS) is one of intriguing objects composed of the densest and most extreme matter,
which is magnetized with a strong magnetic field. Its giant moment of inertial makes it
be a stable clock-like rotator, spinning down gradually by the emission of
electromagnetic wave, particle wind, etc. Precise timing measurement of pulsar emission
can provide deep insight into the interior structures or evolution of its composition and magnetic field,
even allow directly detection of a stochastic gravitational waves background \cite{Jenet2005}. The timing precision is
affected by two main factors or irregularities, glitching and timing noise. The former is a sudden increase and relaxation
of the rotation rate believed to be caused by a sudden unpinning and repinning of the superfluid vortices inside
NS \cite{Anderson1975,Pines1985,Link1992}, whereas the latter is characterized by a
continuous and unpredictable pulse phase wandering
\cite{Lyne1995}. The amount of this timing noise can be measured from the timing residual,
which is a difference of pulse time of arrival (ToA) between a timing model, including glitching, and the real observation.

A single ToA can be obtained from comparing an average pulse profile of many pluses with a
template profile at observatory. Actual ToAs are affected by secular spin-down and
many correcting factors, including pulsar position,
proper motion, dispersion and its variation, and the Earth motion relative to the solar system barycenter \cite{Edwards2006}.
Thus, after these spin-down and correcting factors are taken into account,
the timing residual most likely comes from the internal torque
fluctuations arising from,  $e.g.$, the pinning-unpinning and creep of the crustal superfluid vortex \cite{Alpar1986},
the magnetosphere activities \cite{Cheng1987}, or the pulsar circumstance such as
the companion or aerolite \cite{Cordes1993,Cordes2008,Shannon2013}.
For most pulsars, the residuals from perfectly timing model should include at least two components: a ``slow" irregularities
(due to glitching and timing noise) and a ``fast" white noise. The slow timing noise is of time-correlated
with a red power spectral, containing various contributions from the examples of intrinsic spin noise and magnetospheric
torque variations et al. \cite{Shannon2014}, while the white noise contain both measurement (radiometer)
noise and the jitter effects from changes of the pulse-to-pulse phase and amplitude \cite{Coles2011}.

Various observational and data analysis aspects of timing noise have been accounted for many pulsars.
An earlier attempt was made to describe power spectra in the red noise as random walks in pulse phase,
spin frequency $\nu$ or frequency derivative $\dot{\nu}$ \cite{Boynton1972,Groth1975ApJS,Cordes1980},
for example, the Crab pulsar was consistent with random walks in frequency.
However, the idealized, large rate random walks
in one or more of the three observation variables is too simple.
Instead, superimposed on the random walks, discrete identifiable ``micro-jumps" events
were proposed in many residual data \cite{Cordes1985,D'Alessandro1995,Lyne2010}.
These microjump amplitudes are still too large to be produced by mere fluctuations of a random walk process;
being not simply scaled-down versions of glitches as well.
(unlike glitches, microjumps exhibit both positive and negative signs in frequency and frequency derivative.)

Following steadily accumulating data base,
many empirical timing residual characteristics of the entire set of pulsars \cite{Cordes1985,D'Alessandro1995,Hobbs2010,Shabanova2013}
have been established.
For example, lots of data observed with the Lovell Telescope over several decades presented the
detailed timing residuals and analysis of 366 pulsars \cite{Hobbs2010}.
The statistical nature of timing activity in these pulsar samples can be summarized as follows:
\begin{itemize}

\item  With approximately 37 per cent of the samples the residuals are dominated by both
measurement and jitter errors. More pulsars (63 per cent) exhibit the
low-frequency structures which have a red power spectrum.
Root mean square (rms) of the residuals ranges from a few milliseconds to several seconds.

\item More residuals (36 per cent) are subject to long-term cubic polynomial components
related to the second frequency derivative $\ddot{\nu}$ or the braking index, $n=\nu\ddot{\nu}/{\dot{\nu}}^{2}$.
The braking indices ranging from 36246 to -287986 for non-recycled pulsars
deviate a value of 3, which was expected for the magnetic dipole radiation.
In these significant $\ddot{\nu}$, almost equal number pulsars have both positive and negative signs.

\item The timing noise strength is only weakly correlated with pulse period, but
strongly correlated with the frequency derivative. Therefore, the noise strength
is strongly anti-correlated with the spin-down or characteristic age ($\tau_{\rm c}=-\nu/2\dot{\nu}$)
and with surface magnetic field ($B_{\rm s}=3.2\times 10^{19}\sqrt{-\dot{\nu}/\nu^{3}}$ G),
which could imply an evolutionary character \cite{Shabanova2013}.

\item Another common characteristic feature of the residuals,
what we have been particularly concerned, shows periodic or quasi-periodic phase variations
with long time scales (about 1 to 10 year).
Table 1 shows some samples of pulsars in references \cite {Hobbs2010,Lyne2010}
with significant periodic structures. It is natural
to expect more candidates if the residuals become
stationary over a very long observational span.

\end{itemize}

Whilst these long time scale structures have been considerable accumulated,
the mechanisms are still not fully understood.
Generally, the timing noise can be ascribed to fluctuations of the moment of inertia \cite{Pines1972},
torque acting on the crust from the internal \cite{Packard1972,Anderson1975,Lamb1978,Alpar1986}
or external action associated with the accretion flow \cite{Lamb1978} and magnetosphere current flow \cite{Cheng1987}.
Theoretical models of these mechanisms are most involved with a random origin of the timing noise.
However, recent efforts were concentrated on causes of non-stochastic timing activity
prompted by the periodic or quasi-periodic feature in timing residuals.
Although the low frequency timing noise structure, such as the cubic term or discrete events
in the residual series, can also exhibit a pseudo-periodicity,
the period and amplitude are dependent on the data span.
These regular modulations with a long time-scale (about order of several years)
are also impossible to be caused by any dynamic process in the magnetosphere \cite{Cordes2013}.
The expected oscillation can be either external to the magnetosphere
or in the interior of NS, but the detailed physical mechanism behind the non-random process is still uncertain.

The existing modulation mechanisms \cite{D'Allesandro1997} include: (\romannumeral1) the vortex array Tkachenko waves
\cite{Horn1980} occurring in the rotating superfluid;
(\romannumeral 2) free precession \cite{Shaham1977,Stairs2000,Jones2012} if the spin axis was not aligned with the angular
momentum vector; and (\romannumeral 3) the presence of a planetary system outside a pulsar \cite{Wolszczan1994}.
These mechanisms maybe explain the periodicity and other peculiar behaviors of pulsar,
but a unified theory covering all aspects of the periodicity and the accompanying performances
is required. Some important accompanying observational performances have:
(1) correlating changes in the pulse shape of PSR B1828-11 \cite{Stairs2000}
or other pulsars \cite{Lyne2010}; (2) switching sharply between two or several magnetosphere states
which was first reported in intermittent pulsar PSR B1934+21 \cite{Kramer2006}
and other pulsars from analysis \cite{Lyne2010} of the 366 samples;
and (3) appearing of long-period and rich harmonic contents. The long-period is incompatible
with the precession mechanism when considering neutron superfluid vortices pinning either to magnetic flux tubes in pulsar core
or to inner crust nuclei \cite{Link2003}. The sharply switching and its trigger mechanism
are not easy to understand with any signal modulation methods above, in spite of the possibility of
the precession trigging \cite{Jones2012}. Although Zhang et al \cite{Zhang2012} has phenomenologically modeled the modulation
in Hobbs samples \cite{Hobbs2010} as some unknowing oscillations of the magnetic
field, the implied physics has not been well understood,
perhaps being interpreted \cite{Goldreich92,Pons2012} as a consequence of the Hall magnetic cascade process.

Here, we propose another modulation mechanism: Landau quantization
and de Haas-van Alphen (dHvA) magnetic oscillation when degenerate electrons orbiting strong magnetic field.
As is well known, thermodynamic properties (such as the magnetization and susceptibility)
of single crystal metal on the external magnetic field $H$ and low temperature condition,
$kT<\hbar \omega_{\rm c}$ ( $\omega_{\rm c}$ is the cyclotron angular frequency of the electron
orbit), is oscillatory in nature.
The reason is the quantization of electron motion perpendicular
to the field direction. As successive Landau levels sweep
through the Fermi energy, physical properties of electron system are periodic functions
of the external field. In pure metal crystal,
the dHvA effect is extensively studied \cite{Shoenberg84,Condon1966,Gordon2003}.
Generalizing the non-relativistic situation in metal to NS interior
has been considered in \cite{Blandford82,Suh10,Wang2013}.

As we shall see there, dHvA effect readily provides a crude mechanism of explaining
the periodic modulation in pulsar timing residuals. Accompanying the internal magnetic field evolution slowly,
this theory gives an rapid oscillatory magnetic dipolar moment as observing by a remote observer.
For a normal NS, the expected modulation period decreases nearly $10^{-4}$ relative the smooth evolution,
and $10^{-5}$ for the relative change of the dipolar moment. In $\S$2 we give a summary of timing observation
for periodic or quasi periodic pulsars. We briefly review dHvA magnetic oscillatory theory and application
to NS in $\S$3. In $\S$4 we discuss the modulation of magnetized field of NS due to dHvA effect
and compare theoretical predictions and timing observations in In $\S$5.
Finally, we give, in $\S$6, a brief summary and discussion.

\section{Summary of periodic or quasi periodic modulation from relevant observations}

The spin change of a pulsar with time is identified as a departure from regular spin down
including several frequency derivatives. The pulse phase \cite{Cordes1993} $\phi$ at the barycentric arrival
time $t$ ($t\ll \tau_{\rm c}$) is
\begin{equation}\label{equ.phase}
\phi(t)=\phi_{0}+\nu t+\dot{\nu} t^{2}/2+\ddot{\nu}t^{3}/6+\phi_{\rm{TN}}(t)+\phi_{\rm {WN}}\mbox{ ,}
\end{equation}
where $\phi_{0}$ and $\nu$ are the initial phase and spin rate, $\dot{\nu}$ and $\ddot{\nu}$ are the
spin-down coming from the secular magnetosphere torque, $\phi_{\rm{TN}}(t)$ is the ``timing noise''
associated with the time span $t$ and $\phi_{\rm {WN}}$ is the white noise.
Thus pulsar ``timing noise'' structure may be obtained from residuals deviating from
the polynomial and white noise.

Most comprehensive observations and timing noise study have presented in \cite{Hobbs2010},
using data sets of 366 pulsars from Jodrell Bank observatory. Several decade duration of the data sets
is important for timing residual structure study to identify long-period harmonic terms.
In this pulsar sample, more pulsars (63 per cent) exhibit the low-frequency structures.
Among them, at least 6 pulsars (for example PSRs B1540-06, B1642-03, B1818-04, B1826-17, B1828-11
and B2148+63) appear with significantly periodic or quasi-periodic structure.
This subclass was extended to 17 pulsars \cite{Lyne2010} which have been analyzed in more detail by Lyne et al.
Those 17 selected pulsars showed large amounts of timing noise and the large ratio of maximum slow-down rate
to the minimum rate. The following quasi-periodic features emerged.

Significant harmonic variation is displayed in spin-down rate of those pulsars
(see fig.2 or figs.S2 and S3 in reference of \cite{Lyne2010}).
Obviously, in the Lomb-Scargle spectra of fig.S2 and the wavelet spectra
of fig.S2 \cite{Lyne2010} in Lyne et al., some appear one frequency periodic features
(such as B1540-06, B1714-34 and B1818-04), whereas others (such as B0950+08, B1642-03, B1839+09,
B1903+07) show broader, less well-defined peaks. Formally, in the several peak case the power spectra seems likely
forming from several time intervals and then added together to constitute the statistically significant spectra.
This correlation implies that pulsar timing noise and power spectra have an evolutionary character.

Due to the limitation of measuring actually spin-down change, the rate of slowing down seems
apparently to show switching phenomena between several (usually two) discrete states.
Ten of the 17 pulsars \cite{Lune2013} displayed synchronized change of pulse radio
emission properties (profile or flux density) correlated with slowdown rate variations.
This implies a link between changes of magnetospheric currents and slowdown rate.
Up to now, however, there is no understanding of what causes these discrete states
and the multi-year quasi-periodic modulation of the statistical properties of these states.
Within the magnetosphere dynamical system the current may oscillate regularly,
when it is forced by a regular external forcing arising from the unpinning of superfluid vortices,
magnetic field variation interior of the pulsar or star quake. Here we only concern the origin of
magnetic field variation.

The basic timing properties of 12 pulsars selected from the 17 objects \cite{Lyne2010}
(except 4 pulsars with maximum timing noise excess one second) in Lyne et al. are summarised
in Table~\ref{tab:table5}, in increasing order of the magnitude of the timing noise.

\begin{table}
\caption{\label{tab:table5}The basic timing properties for some
pulsars with periodicity or quasi-periodicity \cite{Hobbs2010,Lyne2010}. In column order,
there are the pulsar name, spin-frequency, frequency derivative,
surface dipole magnetic field, characteristic age, major period,
and residual range (actually taking the difference of the maximum and minimum residuals)
%These data come from
 .}
\begin{ruledtabular}
\begin{tabular}{ccccccc}
 PSR  &$\nu$ &$\dot{\nu}$ &$B_{s}$
 &$\tau_{c}$  &$\tau$ & Residual \\
 &(s$^{-1}$) &($10^{-15}$s$^{-2}$)&(10$^{12}$G)&(M yr)
 &(yr)&(ms)\\
\hline
B2148+63&2.63 &-1.18 &0.3 &35.3
& 3.2 & 3.6\\
B0950+08&3.95 &-3.59 &0.2 &17.4
& 14.3 & 16.1\\
B1540$-$06&1.41 &-1.75 &0.8 &12.8
& 4.2 & 36.3\\
B1826$-$17&3.26 &-58.85 &1.3 &0.9
& 3.0&47.1 \\
B1714$-$34&1.52 &-22.75 &2.6 &1.1
& 3.8 & 64.8\\
B1642$-$03&2.58 &-11.85 &0.8 &3.5
& 6.6 & 83.6\\
%B1133+16&0.84 &-2.65 &2.1 &5.0
%& 28.0& 39.6\\
%B0329+54&1.40 &-4.01 &1.2 &5.5
%&16.9 & 39.9\\
%B2217+47&1.86 &-9.54 &1.2 &3.1
%&14.0 & 45.0\\
\\
%B1826$-$17&3.26 &-58.85 &1.3 &0.9
%& 3.0&47.1 \\
%B1541+09&1.34 &-0.77 &0.6 &27.6
%& 7.0 & 80.9 \\
%B1642$-$03&2.58 &-11.85 &0.8 &3.5
%& 6.6 & 83.6\\
B1907+00&0.98 &-5.33 &2.4 &2.9
& 6.7 & 93.0\\
B1929+20&3.73 &-58.63 &1.1 &1.0
&1.7 & 109.7 \\
B1839+09&2.62 &-7.50 &0.7 &5.5
& 1.0 & 142.5\\
B1818$-$04&1.67 &-17.70 &1.9 &1.5
& 9.1& 595.1\\
B0919+06&2.32 &-73.98 &2.4 &0.5
&1.6 & 708.3 \\
B0740$-$28&6.00 &-604.36 &1.7 &0.2
& 0.4 & 813.0\\
%1828$-$11&2.47 &-365.84 &4.9 &0.1
%& 1.4& 8.9$-$30.5 \footnotemark[1]\\
\end{tabular}
\end{ruledtabular}
%\footnotetext[1]{Pulsar timing residuals after the removal of a cubic term.}
%\footnotetext[2]{Here's the second.}
%\footnotetext[3]{Here's the third.}
%\footnotetext[4]{Here's the fourth.}
%\footnotetext[5]{And etc.}
\end{table}

\section{preliminary theory of dHvA oscillation in neutron star interior}

A comprehensive and general discussion of dHvA magnetic oscillatory has been given in \cite{Shoenberg84}.
Its astrophysical applications, especially in NS, was done
by many authors, including R D Blandford and L Hernquist \cite{Blandford82}.
For completeness, we summarise these discussions and results.

\subsection {De Haas-van Alphen magnetic oscillatory theory}

Considering a uniform magnetic field $H$ directed along the z-axis, transverse
motion of electron is quantized with the condition of area quantization,
$a(\varepsilon,k_z)=(n+1/2)2\pi e H/\hbar c$. Here $n=0,1,2,\ldots$,
$a$ is the area of cross-section cut by a plane normal to
$H$ in wave vector space with constant energy surface of $\varepsilon$,
and $k_{ z}$ is the wave vector component along the z-direction.
The Landau energy levels are obtained by solving the Dirac equation \cite{Jonhson49}
\begin{equation}\label{equ.level}
\varepsilon=[c^{2}(\hbar k_{z})^{2}+\varepsilon_{\rm e}^{2}+\varepsilon_{\rm e} \varepsilon_{\rm c}(2n+s+1)]^{1/2}\mbox{ ,}
\end{equation}
where $\varepsilon_{\rm e}\equiv m_{0}c^{2}$ and $\varepsilon_{\rm c}\equiv \hbar eH/cm_{0}$ are the rest energy and cyclotron energy of
electron respectively, $s=\pm 1$ is the spin quantum number.
The number of states per interval $\mathrm {d}k_{z}$ with a given spin $s$ and Landau quantum numbers $n$
in a volume $V$ must be $V d k_{z}eH/(4\pi^{2}\hbar c)$.

Theoretical outlook of the dHvA effect is summarized here with reference to
Shoenberg \cite{Shoenberg84}. For a system of degenerate Landau levels specified by Eq.(\ref{equ.level}),
traditional approach of derivations of thermodynamic quantities is
to calculate the grand potential
\begin{equation}\label{equ.grand}
\Omega=-kT\int_{-\infty} ^{\infty} \mathrm{d}k_{z}\left(\frac{eHV}{4\pi^{2}c\hbar}\right)
\sum_{n,s}\ln(1+e^{(\zeta-\varepsilon)/kT})\mbox{ ,}
\end{equation}
where $\zeta$ is the chemical potential. For the mathematical simplifications
and explicit physical meaning, the calculation of Eq.(\ref{equ.grand}) has been
broken down into some simpler steps as suggested in \cite{Shoenberg84}:
first carrying through the calculation at $T = 0$ and then taking into
account the phase smearing effect. As $T = 0$, the potential $\Omega$ reduces to
\begin{equation}\label{equ.grand-1}
\Omega=\int_{-\infty} ^{\infty} \mathrm{d}k_{z}\left(\frac{eHV}{2\pi^{2}c\hbar}\right)
\sum_{n=0}^{n_{\rm max}}(\varepsilon_{n}-\zeta')\mbox{ .}
\end{equation}
The summation is now only over $n$ ($\varepsilon_{n}<\zeta'$, $\zeta'$ is chemical potential at $T=0$)
and considering, for the moment, the spin is degenerate.
This can be approximated by the Euler-Maclaurin formula.
As far as the oscillatory behaviour is concerned, only the contribution is considered
as $n$ passes through the Fermi level. This would display a periodicity in field $H$.
Integration over $k_{z}$ in Eq.(\ref{equ.grand-1}) is an oscillatory integral
and the result comes from the phase stationary points, i.e. the extremal cross-section area of Fremi surface.
Thus, the oscillatory part of the grand potential becomes
\begin{equation}\label{equ.grand-2}
\widetilde{\Omega}=\left(\frac{e}{2\pi c\hbar}\right)^{3/2}\frac{\beta V H^{5/2}}{\pi^{2}
(|A^{\prime\prime}|)^{1/2}}
\sum_{p=1}^{\infty}\frac{1}{p^{5/2}}\cos[2p\pi(\frac{F}{H}-\frac{1}{2})-\frac{\pi}{4}]
\mbox{,}
\end{equation}
where $\beta\equiv e\hbar/mc$ is the double Bohr magneton with the cyclotron mass
$m\equiv\frac{\hbar^{2}}{2\pi}\left(\frac{\partial A}{\partial\varepsilon_{\rm F}}\right)$,
$A^{\prime\prime}=\left(\frac{\partial^{2} A}{\partial k_{z}^{2}}\right)$,
$A$ is the extremal cross-section area of Fermi surface.
Eq.(\ref{equ.grand-2}) shows that the potential oscillates with fundamental
dHvA frequency $F$ proportional to $A$,
$F=(c\hbar/2\pi e)A$. By differentiating the potential, we can obtain oscillatory first-order
thermodynamic quantities (the pressure, internal energy and entropy et al.)
also the second-order thermodynamic quantities (such as the
susceptibility, the heat capacities et al.).

However, this highly idealized situation must be modified by finite temperature, electron's collisions
and spin. In effect, finite temperature blurs out the completely sharp Fermi surface
at $T=0$; electron collisions should broaden the fine Landau levels;
electron spin is thought of having separate Landau levels with spin-up and spin-down.
These effects can be contemplated by the so called ``phase smearing"
that equivalently superposes the oscillations of Eq.(\ref{equ.grand-2}) with various frequency $F$,
or equivalently the phase over a small range around the idealized situation.
The phase smearing will reduce the oscillatory amplitude and there is a phase shift.
Based on the theory of phase smearing, discussed in detail by shoenberg \cite{Shoenberg84},
the effects of finite temperature, electron scattering and spin are to multiply amplitude respectively
by the reduction factors of $R_{T}$, $R_{\rm D}$ and $R_{\rm S}$,
and there is no phase shift.
For the $p$th harmonic term, the temperature reduction factor is
\begin{equation}\label{equ.T-factor}
R_{T}(p)=\frac{\pi\lambda}{\sinh\pi\lambda}=\frac{2\pi^{2}pkT/\beta H}{\sinh(2\pi^{2}pkT/\beta H)}\mbox{.}
\end{equation}
For electron collision or scattering by particle or quasi-particle with a
finite relaxation time $\tau$, the reduction factor \cite{dingle1952} is
\begin{equation}\label{equ.D-factor}
R_{\rm D}(p)=\rm {e}^{-\pi p\hbar/\beta H \tau}=\rm{e}^{-\pi p/\omega_{\rm c}\tau}\mbox{,}
\end{equation}
where the cyclotron angular frequency $\omega_{\rm c}$ is defined as
$\omega_{\rm c}\equiv \frac{2\pi e H}{c\hbar^{2}}/\left(\frac{\partial A}{\partial \xi}\right)_{k_{z}}$.
In a magnetic field the energy level of electron spin is split in two, $\varepsilon\pm \triangle \varepsilon/2$,
according as spin-dow or spin-up. Correspondingly the reduction factor becomes
\begin{equation}\label{equ.S-factor}
R_{\rm S}(p)=\cos(p\pi\triangle\varepsilon/\beta H)\mbox{.}
\end{equation}
Having the preliminary theory preparation common to all the methods of dHvA effect,
application in NS will be discussed below.

\subsection {Magnetic oscillatory in magnetized neutron star}
It was widely accepted that gas of degenerate electrons is prevailing in NS
crust and core as a background of atomic nuclei, protons, neutrons and other
fermions and/or bosons. For a reasonable approximation of free electrons, the Fermi energy
$\varepsilon_{\rm F}$ is calculated from
$\varepsilon_{\rm F}=c\sqrt{(m_{0}c)^2+\hbar^2(3\pi^{2}n_{\rm e})^{2/3}}$, where
$n_{\rm e}$ and $m_0$ are the number density and rest mass of electron respectively.
For mass density $\rho\gg 10^6 $ g $\rm {cm}^{-3}$, the electron is ultra-relativistic and
the Fermi energy is $\varepsilon_{\rm F}\approx 51\rho_{12}^{1/3}Y_{\rm e}^{1/3}$ Mev,
here $Y_{\rm e}$ is the electron fraction and $\rho_{12}$ scaled by $10^{12}$ g $\rm {cm}^{-3}$.

In the interior of NS, the Fermi surface structure for ideal electron gas is approximately
sphere. According to the approximation, the extremal area $A$ of cross-section of the Fermi surface by a plane normal
to $k_{z}$ can be obtained by the relation of
$\varepsilon_{\rm F}^{2}=\varepsilon_{\rm e}^{2}+c^{2}\hbar^{2}A/\pi$,
correspondingly the dHvA frequency becomes
\begin{equation}\label{equ.frequency}
F=\frac{c\hbar}{2\pi e}A=\frac{\varepsilon_{\rm F}^{2}-\varepsilon_{\rm e}^{2}}{2ec\hbar}\mbox{.}
\end{equation}
Similarly, we can calculate the cyclotron mass $m$ and angular frequency $\omega_{\rm c}$ of electron respectively,
$m\equiv\left(\frac{\partial A}{\partial\varepsilon_{\rm F}}\right)\hbar^{2}/2\pi=\varepsilon_{\rm F}/c^2$
(i.e., relativistic mass) and $\omega_{\rm c}\equiv \left(\frac{\partial \varepsilon_{\rm F}}{\partial A}\right) 2\pi eH/c\hbar^{2}
=eH/cm$. Another intuitive physical meaning of the cyclotron angular frequency indicates
that the transition between two successive energy levels equals $\hbar \omega_{\rm c}$.
Also, simple calculations show $|A^{\prime\prime}|=2\pi$.

It was well known that the dHvA effect would occur for $kT\leq\hbar \omega_{\rm c}\ll\varepsilon_{\rm F}$.
Condition of $\hbar \omega_{\rm c}\ll\varepsilon_{\rm F}$ means many Landau levels are occupied by electrons.
As suggested by Shoenberg \cite{Shoenberg84} the actual maximum Landau quantum number is
appropriate as long as $n_{\rm {max}}>1$, corresponding to
$\varepsilon_{\rm F}>\varepsilon_{\rm {FC}}\equiv\varepsilon_{\rm e}\sqrt{1+2\varepsilon_{\rm c}/\varepsilon_{\rm e}}$
(notice the dependence of $\varepsilon_{\rm c}$ with magnetic field $H$).
As $\varepsilon_{\rm F}\leq\varepsilon_{\rm {FC}}$, the electron gas is the ultra-magnetized
and the dHvA effect is negligible. On the contrary, the gas becomes non-magnetized
as $2\pi kT\geq\hbar \omega_{\rm c}$ (or $\beta H$). In this case, the temperature reduction factor of
$R(1)\leq 0.14$ in Eq.(\ref{equ.T-factor}) and the Landau levels are considered as quasi-continuous
which suppress the contribution of magnetic oscillation.
It is suitable to define a critical temperature $T_{\rm c}^{(\rm {osci})}$ separating the regions of non-magnetized
and weak magnetized (or intermediate magnetized),
\begin{equation}\label{equ.cri-1}
kT_{\rm c}^{(\rm {osci})}\equiv \frac{\hbar \omega_{\rm c}}{2\pi}
=\frac{\varepsilon_{\rm e}}{2\pi}\left(\frac{H}{H_{\rm Q}}\right)
\left(\frac{\varepsilon_{\rm F}}{\varepsilon_{\rm e}}\right)^{-1}\mbox{,}
\end{equation}
where $H_{\rm Q}=4.414\times 10^{13}$ G is the relativistic magnetic field.
For fixed $H$ of Eq.(\ref{equ.cri-1}), one obtains the linear $\varepsilon_{\rm F}^{-1}$
dependence of the critical temperature. This indicates that the most likely location of
the dHvA oscillation occurs in the outer crust. For a Fermi energy \cite{Chamel2012} of 4.3 MeV at
$\rho\approx 1.3\times 10^{9}$ g cm$^{-3}$ ($_{28}^{64}$Ni) and $H=H_{\rm Q}$,
for instance, the critical temperature may reach up to $10^{8}$ K,
much higher than the internal temperature \cite{Hernquist1985}
of normal NS, even comparable to the Crab. With a magnetic field up to $10^{15}$ G in magnetar
\cite{Duncan92}, the location may extend deeper to the inner crust or inner core.
However, we will focus on the outer crust of NSs below.
In figure 1, we display the regions of validity of the dHvA oscillation.
\begin{figure}
\centering
  %Requires \usepackage{graphicx}
  \includegraphics[scale=0.35, angle=-90]{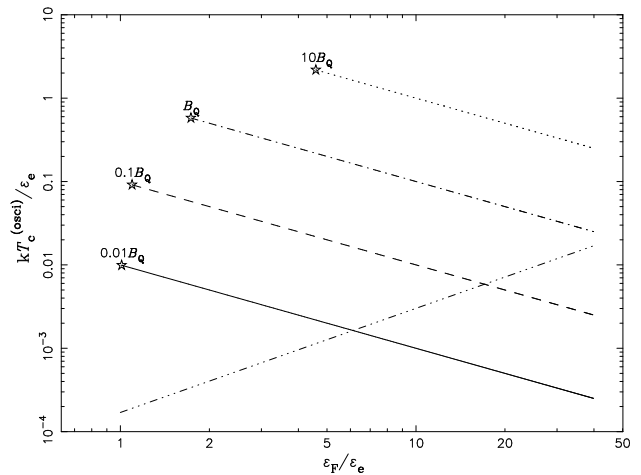}
  \caption{Diagram of electron Fermi energy versus critical temperature for dHvA magnetic oscillatory in
magnetized NS outer crust with magnetic field 0.01 $B_{\rm Q}$ (the solid line),
0.1 $B_{\rm Q}$ (the dash line), $B_{\rm Q}$ (the dot-dash line) and 10 $B_{\rm Q}$ (the dot line).
$B_{\rm Q}=4.42\times10^{13}$ G, $\varepsilon_{\rm e}=0.5$ MeV. The dot dot dot-dash line is the linear
approximation of temperature from the surface ($10^{6}$ k) to inter crust ($10^{8}$ k).}
%  \label{b}}
\end{figure}

The reduction factor of electron scattering stems from phonons and impurities
of the crust. Apart from very young NS, the phonon scattering is negligible \cite{cumming2004}.
The Dingle temperature $T_{\rm D}$, defined as $2\pi\hbar\tau/k$,
due to impurity scattering \cite{Itoh1993} is
\begin{equation}\label{equ.Dingle}
T_{\rm D}\approx 4.0\times 10^{5}\Lambda_{\rm
eQ}\left(\frac{\varepsilon_{\rm F}}{4.3 \rm {Mev}}\right)\left(\frac{Z}{28}\right)^{-1}
\left(\frac{Q}{10}\right)\\\ {\rm K},
\end{equation}
where $\Lambda_{\rm eQ}$ is the Coulomb logarithm of order
unity, $Z$ is the charge number of the atomic nucleus,
$Q$ is the impurity factor which is
defined as the mean square charge deviation $\langle (\triangle
Z)^{2}\rangle$. Here we have scaled the impurity $Q$ \cite{Jones2001} by a large
value $Q\sim10$. In spite of such consideration of $Q$, the impurity scatting becomes relatively less important.
The picture discussed above is consistent with very high electric and
thermal conductivities of NS interior.
With a fairly simple treatment, we may note that the reduction factor in Eq.(\ref{equ.S-factor})
becomes $(-1)^{p}$, corresponding to a spin-lifting equals to the Landau level spacing.

Other thermodynamics and kinetic quantities of NS matter can be derived from the grand potential.
First-order thermodynamical quantities, such as the pressure, the entropy and the magnetic moment
(or magnetization), and the second-order quantities (heat capacities, magnetic susceptibility) can
be obtained by differentiating the potential one or twice. The oscillatory magnetization $\widetilde{M}$ follows
by differentiating the potential density with respect to $H$. In this process,
only the relatively rapidly varying sinusoidal factor in Eq.(\ref{equ.grand-2})
is valid at usual approximation of differentiating.
After introduction of two reduction factors in Eq.(\ref{equ.T-factor}) and Eq.(\ref{equ.S-factor}),
the oscillatory magnetization is
\begin{equation}\label{equ.magnatization-1}
\widetilde{M}=-M_{0}(\varepsilon_{\rm F},H)
\sum_{p=1}^{\infty}\frac{(-1)^{p}R_{T}(p)}{p^{3/2}}\sin[2p\pi(\frac{F}{H}-\frac{1}{2})-\frac{\pi}{4}]
\mbox{,}
\end{equation}
and, for ultra relativistic Fermi energy, the major amplitude of oscillation can be expressed as
\begin{equation}\label{equ.amplitude-0}
M_{0}(\varepsilon_{\rm F},H)=\frac{\alpha}{4\pi^{3}}\left(\frac{\varepsilon_{\rm F}}{\varepsilon_{\rm e}}\right)
\left(\frac{H}{H_{\rm Q}} \right)^{-1/2}H\mbox{,}
\end{equation}
where $\alpha$ is the fine structure constant, the temperature reduction factor of $ R_{T}(p)$
is included in the phase smearing.
If taking the Fermi energy as several MeVs and the field as $H_{\rm Q}$,
the relative magnetization to the external field, $4\pi M_{0}/H$,
is about $10^{-3}$,  which is not negligible.
In practice, the oscillation phase appearing in Eq.(\ref{equ.amplitude-0}) is equivalent to
another kind of, $2\pi(\frac{F}{H}-\frac{1}{2})-\frac{\pi}{4}=2\pi kh$.
Here, $h=H-H_{0}$ is the change of the field $H$ relative to $H_{0}$ which is the
oscillatory centre at which the magnetization vanishes,
$k=1/2(\varepsilon_{\rm F}/\varepsilon_{\rm e} )^2 (H_{0}/H_{\rm Q})^{-1}H_{0}^{-1}$
(as $\varepsilon_{\rm F}\gg \varepsilon_{\rm e}$).

The Lifshitz-Kosevich theory above is linear dHvA effect in that electron only feels
magnetic field $H$, neglecting the feedback contribution from the magnetization $\widetilde{M}$.
Correct procedure allowing for the so called magnetic interaction is only to replace $H$ by
magnetic induction $B$ in Eq.(\ref{equ.magnatization-1}). An important implication of the no-linear
dHvA oscillation is responsible for magnetic instability of the electron gas.
Only those states under the condition, $\partial B/\partial H>0$, are thermodynamically stable.
With strong dHvA effect, i.e. the differential magnetic susceptibility
$\chi_{\rm m}=\partial(4\pi M)/\partial H>1$, homogeneous magnetized state transforms into
Condon domain structure \cite{Condon1968}. The instability condition and feature of magnetic domains and domain walls
were studied in more detail in \cite{Gordon2003} and \cite{Wang2016}.

\section{de Haas-van Alphen oscillatory Modulation of pulsar magnetic dipolar moment}

Possible observational consequences of the dHvA effect and magnetic phase transition
initiate a great interest in compact stars. Generally, higher-order thermodynamic quantities
suffer much stronger affection of magnetic field than the first-order quantities.
Hence fraction variation of the differential magnetic susceptibility is larger than the
fraction variation of magnetization. For NS spinning-down by magnetic dipole radiation,
this means a larger fraction variation of second frequency derivative $\ddot{\nu}$
or larger braking index deviation from $n=3$. Either variation of the field or the Fermi energy
can introduce the dHvA oscillation. For the pulsar timing analysis,
dynamic magnetized process of varying the field with time will be particularly concerned.

The overall evolution of the NS magnetic field is completely dominated by the Ohmic dissipation,
ambipolar diffusion and Hall drift. The dependence of relaxation time scales $\tau_{\rm m}$
(up to $10^{4}$ years or even more large) on the Hall effect and Ohmic decay in NS crust
is a complicated problem.
In an observed time span $t$ which is short relative to the time scale $\tau_{\rm m}$,
a slow evolution means that the interior field is quasi-magnetostatic.
This implies that the spatial change can be separated from the temporal one,
$H(r,t)=H_{0}(r)(1-t/\tau_{\rm m})$. Correspondingly, the fundamental time frequency of the dHvA oscillation is given by
$\nu=kH_{0}(r)/\tau_{\rm m}$, which is speeded up by a factor of $kH_{0}(r)$.
Substituting the formula of $k$ given above for the time frequency, we obtain
\begin{equation}\label{equ.frequency-1}
\nu\tau_{\rm m}\approx 1.6\times 10^{3}\left(\frac{\varepsilon_{\rm F}}{4.3 \rm{Mev}} \right)^2
\left(\frac{H_{0}}{10^{12}\rm{G}}\right)^{-1}\mbox{.}
\end{equation}

The total oscillatory magnetic dipole moment in the outer crust of NS
is a possible coherent sum of many sinusoidal contents.
Assuming an axially symmetric configuration of the magnetic field and matter,
only the component along the axial direction ($z$ direction) is interested for a remote observer.
Non-oscillatory magnetic dipole moment originating from conduction currents
may be expressed as $\overline{m}_{\rm z}=\frac{3}{4}\times (10^{12}\mathrm{G})R_{\rm \star}^{3}\hat{m}$,
where $\hat{m}=\int_{0}^{1}dr
\int_{-1}^{1}(\frac{H_{0}}{10^{12}\rm G})r^{2}\cos\gamma dx$
is a dipolar moment scaled by a normal NS with a homogeneous field $10^{12}$ G.
In the same way, $r$ is the radius of spherical shell scaling in the NS radius $R_{\star}$.
The function $x$ is defined as $x= \cos\theta$, where $\theta$ is the polar angle and $\gamma$ is the
inclination angle between the local magnetic field and axial direction.
%{\Large\bf (Please see if this makes sense)}
%between the directions of the magnetic axis and the local field.
The oscillating part (only considering the fundamental content) of the dipole moment component $\widetilde{m}_{\rm z}$
scaling in $\overline{m}_{\rm z}$ is a double integrals of Eq.(\ref{equ.magnatization-1}),
\begin{equation}\label{equ.moment}
 \frac{\widetilde{m}_{\rm z}}{\overline{m}_{\rm z}}=\texttt{Re}\int_{0}^{1}\mathrm dr
 \int_{-1}^{1}M_{\rm z}(r,x)e^{i\xi \phi(r,x)}\ \mathrm dx,
\end{equation}
with a phase factor $\xi=10^{4}t/\tau_{\rm m}$ and phase function
$\phi(r,x)=(\frac{\varepsilon_{\rm F}}{4.3 \rm MeV})^{2}(\frac{H_{0}}{10^{12}\rm G})^{-1}$.
The relative oscillatory amplitude is determined by
\begin{equation}\label{equ.amplitude}
  M_{\rm z}=1.5\times 10^{-2} R_{T}(1)\left(\frac{\varepsilon_{\rm F}}{4.3 \rm MeV}\right)\left(\frac{H_{0}}{10^{12}\rm G}\right)^{\frac{1}{2}}
  \left(\frac{r^{2}\cos\gamma}
    {\hat{m}}\right).
\end{equation}
Here, we had scaled the Fermi energy with 4.3 MeV at the outer crust of
$\rho\approx 1.3\times 10^{9}$ g cm$^{-3}$
($_{28}^{64}$Ni) and the field with $10^{12}$ G.
This is not extremely sensitive to a select of the characteristic place,
as the Fermi energy increases and the magnetic strength decreases, as approaching the interior of NS.

Because of the large phase factor $\xi$ (with a long observation time span), even a small change of the phase function $\phi$
will generate rapid oscillations in the integral of the magnetization,
which leads to incoherent cancellations. We can use the stationary phase approximation or
the steepest descents method to evaluate the integral as done by Brillouin and Sommerfeld.
A detailed and rigorous account in asymptotic theory can find from the book of Wong \cite{Wong2001}.
The resulting asymptotic expansion comes only from certain critical points
(including the stationary ones) in the phase function $\phi$.
Here we mainly consider the interior stationary points which are non-degenerate,
often called critical points of the first kind.
Other uniform treatments of the critical points on the boundary of the integral or
degenerate stationary points have been given in  \cite{Wong2001}.
Let the real place point of $(r_{n},x_{n})$ be the $n-$th stationary point, the initial terms of $\phi$
in the expansion of Maclaurin have the form of $\phi=\phi_{n}+a_{n}r^{2}+2b_{n}rx+c_{n}x^{2}$.
The leading term in the asymptotic expansion of the oscillatory magnetic dipole moment
at the critical point yields
\begin{equation}\label{equ.moment-1}
 \frac{\widetilde{m}_{n}(\xi)}{\overline{m}_{\rm z}}=\texttt{Re}
 \frac{\pm 2\pi i}{\xi}M_{n}\triangle_{\rm n}^{-\frac{1}{2}}e^{i\xi\phi_{ n}} \mbox{,}(\triangle_{n}>0),
\end{equation}
or
\begin{equation}\label{equ.moment-2}
 \frac{\widetilde{m}_{n}(\xi)}{\overline{m}_{\rm z}}=\texttt{Re}
 \frac{ \pi}{\xi}M_{n}|\triangle_{n}|^{-\frac{1}{2}}e^{i\xi\phi_{n}} \mbox{,}(\triangle_{n}<0),
\end{equation}
where the discriminant is given by $\triangle_{n}=a_{n}c_{n}-b_{n}^{2}$,
and the upper sign in Eq.(\ref{equ.moment-1}) corresponds to a local minimum, while
the lower one to a local maximum.
The total oscillatory magnetic dipole moment $\widetilde{m}_{\rm z}$ should include all oscillate terms at
every critical point. During a short observational period $t$ ($t\ll \tau_{\rm m}$),
the moment can be written formally as
\begin{equation}\label{equ.moment-3}
\frac{\widetilde{m}_{\rm z}}{\overline{m}_{\rm z}}= 0.9\times10^{-5}
 \sum_{n}a_{n}\sin(\omega_{n}t+\varphi_{n}).
\end{equation}
%\begin{multline}\label{equ.moment-3}
%  \frac{\widetilde{m}_{\rm z}}{\overline{m}_{\rm z}}= 0.9\times10^{-5}\times\{
% \sum_{\rm l}a_{\rm l}^{(\rm l)}sin(\omega_{\rm l}t+\varphi_{1})+ \\
%  \sum_{\rm m}a_{\rm m}^{(\rm 2)}sin(\omega_{\rm m}t+\varphi_{2})+
%\sum_{\rm n}a_{\rm n}^{(\rm 3)}sin(\omega_{\rm n}t+\varphi_{3})\}
%\%end{multline}
Here, at the $n$-th critical point, the initial phase is $\varphi_{n}$, and the oscillatory angle frequency is
\begin{equation}\label{equ.frequency-2}
 \omega_{n}=10^{4}\left(\frac{\varepsilon_{\rm F}}{4.3 \rm MeV}\right)^{2}\left(\frac{H_{0}}{10^{12}\rm G}\right)^{-1}\frac{1}{\tau_{\rm m}}.
\end{equation}
The corresponding amplitude at the extreme stationary point ($\triangle_{n}>0$) is
\begin{equation}\label{equ.amplitude-1}
a_{n}=R_{T}(1)\left(\frac{\varepsilon_{\rm F}}{4.3 \rm MeV}\right)\left(\frac{H_{0}}{10^{12}\rm G}\right)^{\frac{1}{2}}
\left(\frac{r^{2}cos\gamma} {\hat{m}}\right)\triangle_{n}^{-1/2} \mbox{ ,}
\end{equation}
or at the saddle point ($\triangle_{n}<0$), it is
\begin{equation}\label{equ.amplitude-2}
a_{n}=\frac{R_{T}(1)}{2}\left(\frac{\varepsilon_{\rm F}}{4.3 \rm MeV}\right)\left(\frac{H_{0}}{10^{12}\rm G}\right)^{\frac{1}{2}}
\left(\frac{r^{2}cos\gamma}
{\hat{m}}\right)\left|\triangle_{n}\right|^{-1/2}.
\end{equation}
In all the calculations given above, the dHvA effect is supposed linear, i.e.
$B=H+4\pi M\approx H$. In what follows we will discuss the no-linear dHvA effect.

Considering the Shoenberg magnetic interaction between electron orbital motions,
correct procedure is only to replace $H$ by magnetic induction $B$ in Eq.(\ref{equ.magnatization-1}).
One of the major modifications is that of leading to the formation \cite{Condon1966} of the so-called Condon domains,
if the differential magnetic susceptibility, $\chi_{\rm m}=\frac{\partial (4\pi M)}{\partial H}>1$.
This is inevitable as $a=8\pi^{2}kM_{0}$ may be larger than unity.
Substituting $k$ and $M_{0}$ from Eq.(\ref{equ.amplitude-0}) we obtain
\begin{equation}\label{equ.critical}
a=4.4\times 10^{2}R_{T}(1)\left(\frac{\varepsilon_{\rm F}}{4.3 \rm{Mev}} \right)^3
\left(\frac{H_{0}}{10^{12}\rm{G}}\right)^{-3/2}\mbox{.}
\end{equation}
%{\Large\bf (Please see if this statement makes sense)}
Condon domains result from the splitting of electron distribution into different
layers \cite{Egorov2010} with different values of magnetization separating by domain walls (DWs).
Magnetizing process of the domain structure is realized by motion of the DWs with an average magnetization,
$4\pi \overline{M}/H_{0}=h/n=t/n\tau_{\rm m}$ in a medium of the demagnetization coefficient $n$.
This is a linear magnetizing process which should not be considered for the timing noises,
except in the harmonic oscillatory regions of $\chi_{\rm m}<1$.
However, the magnetizing process by motion of the DWs is irreversible due to the energy barriers
arising from a variety of disordering defects inside the crust, such as impurities, dislocations,
grain boundaries, etc. Calculations \cite{Horowitz2009} with a large impurity factor $Q>10$ indicate strong
magnetic hysteresis. In this case, the actual dynamical process of DWs is
a characteristic feature of the depinning transition
which is slowly driven by a changing magnetic field through a disordered medium.
This like-avalanche jerky motion of DWs among the pinned metastable configurations may
produce the resemblance of the Barkhausen noise \cite{Narayan1996}.
A further study and observable effects of the dynamical phase transition will defer to future.
\section{Compares the theory with the timing observations of pulsars}

The detailed oscillatory property depends on a realistic profile of the magnetic field
and the composition,  which  does not yet exist. However,
we can summarize the key arguments favoring the magnetic oscillation.
In order to appreciate the observing
possibilities it is useful to deduce from the orders of magnitude involved in $\widetilde{m}_{\rm z}$.
The following observed application and comparison from the model are performed.

\verb+(1) Modulation period: + %{\em\bf (1) Oscillate period:}
Low-frequency Fourier components in an observed residual series
have a natural explanation which are clearly ascribed to the dHvA magnetic oscillation
at stationary points of the phase function. %{\Large(Please see this makes sense)}
It is expected from Eq.(\ref{equ.frequency-1}) or Eq.(\ref{equ.frequency-2})
that the $n$-th modulation period,
$\tau_{n}=2\pi\times10^{-4}(\frac{\mu_{\rm F}}{4.3 \rm MeV})^{-2}(\frac{H_{0}}{10^{12}\rm G})\tau_{\rm m}$,
decreases nearly $10^{4}$ relative to the long magnetic field evolutionary scale $\tau_{\rm m}$.
A global field supported by conduction currents is evolutionary
through three mechanisms, i.e., the Ohmic decay,
ambipolar diffusion and the Hall drift. It was generally considered
that the Hall drift dominates the field evolution in the crust and core of NS.
Further detail has not been well understood, but the drift should be impossibly linear
and be far more complex due to the turbulent Hall cascade \cite{Goldreich92}
or the Hall drift instability \cite{Rheinhardt2002}.
%{\em which may be impossible linear and even more complex
%due to the turbulent Hall cascade \cite{Goldreich92} or the Hall drift instability}
%\cite{Rheinhardt2002} {\Large\bf (Here the last sentence is not clear. Please rewrite it)}.
Depending on the initial magnetic field strength and its gradient,
this process can lead to nonlocal transfer of energy from large to small scales or from
the poloidal to toroidal components. For a normal NS with a field of
$(10^{11}$-$10^{13})$ G, the Hall drift is on the timescale of $(10^{4}$-$10^{6})$ years
\cite{Pons2010}, which implies the oscillatory period of the range ($1-10^{2}$)
years,  close to  the observed modulation period (See Table~\ref{tab:table5}). The puzzle of
the Fourier component \cite{Stairs2000} of 1000 days in the residuals of PSR $1828-11$
would be solved, if the change in the dipole moment was caused by the dHvA oscillation at several stationary points
instead of precessions.
If it is the Hall drift that dominates the field evolution, and the time scale, $\tau_{\rm m}\propto H_{0}^{-1}$,
Eq.(\ref{equ.frequency-2}) implies the oscillating frequency dependence only on the Fermi energy
of the stationary point. This indicates that  the higher frequency oscillation should occur
at deeper interiors of pulsar.
Dispersion of the oscillation periods in Table~\ref{tab:table5}
manifests the indeterminacy of the stationary point distribution.
The change of stationary points in single pulsar should correlate with the Hall cascade.

\verb+(2) Braking index:  + Another important observable quantity in the pulsar timing is the braking index.
From a general thermodynamic theory, higher-order thermodynamic quantities
suffer much stronger affection of magnetic field change than the first-order quantities.
This means a larger fraction variation of second frequency derivative $\ddot{\nu}$
or larger braking index deviation from $n=3$ of constant dipolar moment.
With a varying magnetic dipolar moment the braking index will be
$n=3+2\left(\frac{\dot{\widetilde{m}}_{\rm z}}{\overline{m}_{\rm z}}\right)\frac{\nu}{\dot{\nu}}
$. The expected range is from $\left(3 - 2\frac{\tau_{\rm c}}{\tau_{n}}\right)$ to
$\left(3+2\frac{\tau_{\rm c}}{\tau_{n}}\right)$. If taking the characteristic age $\tau_{\rm c}$ as $10^{6}$ years,
oscillatory timescale $\tau_{n}$ as 10 years (means $\tau_{\rm m}\sim 10^{4}$ years),
for example, then the index is from $-10^{5}$ to $10^{5}$. This  agrees  with
the observation values ranging from $-287 986$ to $+36 246$, with almost equal numbers
of positive and negative signs in the Hobbs sample \cite{Hobbs2010}.

\verb+(3) Residual range: + For a pulsar,  from Eqs.(\ref{equ.moment-3}),(\ref{equ.amplitude-1}) and (\ref{equ.amplitude-2}), we
can  approximatively  obtain the residual range from the minimum to the maximum
(actually taking their difference).
An fraction of oscillatory dipolar moment $\widetilde{m}_{\rm z}/\bar{m}_{\rm z}$
are associated directly with the relative change
of the first derivative of frequency $\delta\dot{\nu}/\dot{\nu}$.
From Eq.(\ref{equ.moment-3}) the maximum variation of $\dot{\nu}$ in a oscillatory cycle is limited to
$(\delta \dot{\nu})_{\rm max}=1.8\times10^{-5}a_{n}\dot{\nu}$.
During an observational period $t$ (near a cycle), the maximum range of
the residuals is given roughly by
%$(\delta t_{\rm R})_{\rm max}=0.9\times10^{-5}\dot{\nu}a_{n}t^{2}/\nu$.
\begin{equation}\label{equ.residual}
\begin{aligned}
(\delta t_{\rm R})_{\rm max}&=142 \rm{\mu s} \left(\frac{\varepsilon_{\rm F}}{4.3 \rm MeV}\right)
\left(\frac{H_{0}}{10^{12}\rm G}\right)^{\frac{1}{2}}\left( \frac{\tau_{\rm c}}{10^{6}\rm{yr}}\right)^{-1} \\
&\times\left(\frac{r^{2}cos\gamma} {\hat{m}}\right)|\triangle_{n}|^{-1/2} R_{T}(1)t_{\rm {yr}}^2 \mbox{.}
\end{aligned}
\end{equation}
In this equation, we have scaled characteristic age $\tau_{\rm c}$ and an observational period $t$ as $10^6$ years
and 1 year respectively.
Like discrete oscillation period distribution, similar residual dispersion involving
the uncertainty of the stationary point may occur.
However, if taking the number value parameter of the stationary point as
$a_{n}\sim 1$, the major oscillation cycle or the observed time span  as $t\sim 10^{9}$ s, the frequency
and its first derivative range as $(1.0-6.0)$ $\mathrm{s}^{-1}$ and $(1.2-604.4)\times10^{-15}$ $\mathrm{s}^{-2}$
respectively, as in Table~\ref{tab:table5}, we find that
the residual range is from the minimal value of 4 ms (PSR B2148+63, real residual 3.6 ms)
to the maximal vale of 906 ms (PSR B0740-28, real residual 813 ms),
which is also in a good agreement with the observation data of pulsars in Table~\ref{tab:table5}.
On the other hand, the predicted tendency of residuals, which  increase with the magnetic field,
also coincides with the data presented in Table~\ref{tab:table5}.

\section{summary and discussion}
 	
We have presented a physical mechanism for the low frequency noises in pulsar
timing observations. In this scenario, the noise is caused by the so-called de Haas-van Alphen effect
originating from Landau quantization of the generate relativistic electron gas.
The long-term time evolution of the magnetic field, usually subject to the Hall drift,
is modulated by the magnetic rapidly oscillation. Because of the relatively poor phase coherence of
local magnetic moment inside a pulsar, non-canceled terms come from stationary
points, including the extreme or saddle points in the phase function. Hence,
Fourier components of timing residuals in pulsars may be ascribed to
the special oscillation. For normal pulsars, the predicted oscillation timescale
(period or cycle), which decreases by a factor of a few thousands or even ten thousands
relative to the field evolution, is consistent with observations. Theoretical amplitude
of the oscillatory magnetic dipole moment is of dependence on the field, Fermi energy,
and their second derivatives at the stationary points. Another overwhelming fact of the theoretical studies,
the braking index corresponding to the estimation of oscillatory amplitude is from $-10^{5}$ to $10^{5}$,
which is also in a good agreement with the observations. For the residual estimation,
there is also consistency between the theory and observation.
Meanwhile, this scenario also indicates a tendency of the noise increase
with the growing surface magnetic field. In some ways, our picture of low-frequency noise is not noise
but reflects the dynamics of magnetic field evolution.
%{\em a tendency of the noise variation
%with the field.} {\Large\bf (Here the last sentence is not clear. Please rewrite it and states clearly what is the tendency.)}

The very discovery of other special class of NSs was possibly associated with
NSs operating in a special evolution phase of the magnetic field.
This special subclass of NS family or manifestations includes, (1) anomalous X-ray (AXPs)
and soft gamma-ray repeaters(SGRs) powered by magnetic field (magnetars)
and low magnetic field like-magnetars;
(2) nulling pulsars with a broad distribution of timescales from fast nulling
with duration of hours or less, the long-term intermittent pulsar on months or years
to the more extreme rotating radio transients (RRATs);
(3) X-ray dim isolated NSs (XDINs) and compact central objects in SNRs (CCOs)
with a common radio-quiet feature. While the ambitious goal to build a
unified electromagnetic radiation mechanism for all the special classes and observational manifestations,
result is still far from being reached to date. The dHvA magnetic oscillatory theory
presented above might provide a possible clue.
In the theoretical model of rapid dHvA modulation of long-term magnetic field evolution,
an indirect evidence is confirmed by direct measurements (e.g. braking index)
of the timing residuals in pulsars. Other dHvA oscillatory properties e.g. the
magnetothermal effect, the magnetostriction effect, the velocity of sound, the electrical properties
have no more special observational consequences.

However, if $\chi_{\rm m}=\frac{\partial (4\pi M)}{\partial H}>1$, the nonlinear effects
due to the magnetic interaction of local dipolar or the Shoenberg effect should
be taken into account. As we have discussed earlier, the stable state of the
electron gas will transform into Condon domains separated by walls.
The magnetization process is most probably made by DW motion
with a dynamics of intimately relating to the wall elastic tension, random quenched
disorder (due to the crust impurities, lattice
dislocations, residual stresses, etc.) and the driving magnetic field of slowly varying.
With increasing or decreasing of the driving field, a sharp threshold field $H_{\rm c}$
will be passing through, below which (i.e. $H<H_{\rm c}$) the wall is pinned by the disorder in one of
many metastable states, and above that ($H>H_{\rm c}$) the wall moves forward with an average velocity.
At the critical field ($H=H_{\rm c}$) of the depinning transition,
the wall moves by avalanches exhibiting scaling invariance
whose sizes and duration distributions follow power laws.
The exponents \cite{Cizeau1997} of the power laws distributions depend on the field driving rate.
We conjecture that this avalanches can trigger a NS ``starquake" of having been invoked
to explain the flashes of electromagnetic radiation from magnetars or like-magnetars even
giant radio pulse of a pulsar. Distinguishing from rotation powered, the activity is powered by
magnetic field with energy deposited in domain structure and the shear deformation of solid crust
(if the depinning transition induces brittle fracture).
Available magnetic free energy  \cite{Wang2016} may reach up to
$10^{38}(B/B_{\rm Q})^{4}$ erg only in domain structure.
The seismic waves however cannot propagate directly to the surface but by many reflections \cite{Blaes1989}
between the surface and evanescent zone (where the magnetic stress begins to dominate
the crust stress). For the torsional wave mode, shaking of the stellar surface twists the magnetic field
anchored in the crust, quite similar to twisting magnetic field lines that extend
beyond light cylinder by stellar rotation. Twisting the open field lines
is responsible for many magnetosphere activities including intensity
nulling, pulse-shape mode changes, subpulse drift rates, spin-down rates,
or as another driving force of the forced Markov processes \cite{Cordes2013}.
In addition, the current induced by the perturbed magnetic field in the crust results in
Joule heating which can be treated as another heat reservoir of XDINs or CCOs.

The main conclusion is that the dHvA magnetic oscillatory and possible impact on NSs
observation are common at least in non young NS crusts in spite of their complex
nature. The complexity includes the coexistence and evolution properties of various harmonic oscillations.
The current scenario is partially able to predict or remove a part of the timing residuals,
which is very important for the background gravitational wave
(overlapping the magnetic oscillatory frequencies) detections
by pulsar timing array.
From a comparison between the expected and observed
oscillatory period and amplitude, we can explore the magnetic field and the composition
of NS interior or their correlations.
%{\Large\bf (Please check the last paragraph, and see if the statements are accurate enough)}

This work is in part supported by the National
Natural Science Foundation of China (Nos. 11563008, 11473024, 11363005, 11463005, 11763007,
11375153 and 11675145)
and XinJiang Science Fund for Distinguished Young Scholars under No. 2014721015.

%\newpage %Just because of unusual number of tables stacked at end
\bibliography{apssamp}% Produces the bibliography via BibTeX.

\begin{thebibliography}{61}
\expandafter\ifx\csname natexlab\endcsname\relax\def\natexlab#1{#1}\fi
\expandafter\ifx\csname bibnamefont\endcsname\relax
  \def\bibnamefont#1{#1}\fi
\expandafter\ifx\csname bibfnamefont\endcsname\relax
  \def\bibfnamefont#1{#1}\fi
\expandafter\ifx\csname citenamefont\endcsname\relax
  \def\citenamefont#1{#1}\fi
\expandafter\ifx\csname url\endcsname\relax
  \def\url#1{\texttt{#1}}\fi
\expandafter\ifx\csname urlprefix\endcsname\relax\def\urlprefix{URL }\fi
\providecommand{\bibinfo}[2]{#2}
\providecommand{\eprint}[2][]{\url{#2}}

\bibitem[{\citenamefont{{Jenet} et~al.}(2005)\citenamefont{{Jenet}, {Hobbs},
  {Lee}, and {Manchester}}}]{Jenet2005}
\bibinfo{author}{\bibfnamefont{F.~A.} \bibnamefont{{Jenet}}},
  \bibinfo{author}{\bibfnamefont{G.~B.} \bibnamefont{{Hobbs}}},
  \bibinfo{author}{\bibfnamefont{K.~J.} \bibnamefont{{Lee}}}, \bibnamefont{and}
  \bibinfo{author}{\bibfnamefont{R.~N.} \bibnamefont{{Manchester}}},
  \bibinfo{journal}{Astrophys. J} \textbf{\bibinfo{volume}{625}},
  \bibinfo{pages}{L123} (\bibinfo{year}{2005}), \eprint{astro-ph/0504458}.

\bibitem[{\citenamefont{{Anderson} and {Itoh}}(1975)}]{Anderson1975}
\bibinfo{author}{\bibfnamefont{P.~W.} \bibnamefont{{Anderson}}}
  \bibnamefont{and} \bibinfo{author}{\bibfnamefont{N.}~\bibnamefont{{Itoh}}},
  \bibinfo{journal}{\nat} \textbf{\bibinfo{volume}{256}}, \bibinfo{pages}{25}
  (\bibinfo{year}{1975}).

\bibitem[{\citenamefont{{Pines} and {Alpar}}(1985)}]{Pines1985}
\bibinfo{author}{\bibfnamefont{D.}~\bibnamefont{{Pines}}} \bibnamefont{and}
  \bibinfo{author}{\bibfnamefont{M.~A.} \bibnamefont{{Alpar}}},
  \bibinfo{journal}{\nat} \textbf{\bibinfo{volume}{316}}, \bibinfo{pages}{27}
  (\bibinfo{year}{1985}).

\bibitem[{\citenamefont{{Link} et~al.}(1992)\citenamefont{{Link}, {Epstein},
  and {van Riper}}}]{Link1992}
\bibinfo{author}{\bibfnamefont{B.}~\bibnamefont{{Link}}},
  \bibinfo{author}{\bibfnamefont{R.~I.} \bibnamefont{{Epstein}}},
  \bibnamefont{and} \bibinfo{author}{\bibfnamefont{K.~A.} \bibnamefont{{van
  Riper}}}, \bibinfo{journal}{\nat} \textbf{\bibinfo{volume}{359}},
  \bibinfo{pages}{616} (\bibinfo{year}{1992}).

\bibitem[{\citenamefont{{Lyne} et~al.}(1995)\citenamefont{{Lyne}, {Pritchard},
  and {Shemar}}}]{Lyne1995}
\bibinfo{author}{\bibfnamefont{A.~G.} \bibnamefont{{Lyne}}},
  \bibinfo{author}{\bibfnamefont{R.~S.} \bibnamefont{{Pritchard}}},
  \bibnamefont{and} \bibinfo{author}{\bibfnamefont{S.~L.}
  \bibnamefont{{Shemar}}}, \bibinfo{journal}{Journal of Astrophysics and
  Astronomy} \textbf{\bibinfo{volume}{16}}, \bibinfo{pages}{179}
  (\bibinfo{year}{1995}).

\bibitem[{\citenamefont{{Edwards} et~al.}(2006)\citenamefont{{Edwards},
  {Hobbs}, and {Manchester}}}]{Edwards2006}
\bibinfo{author}{\bibfnamefont{R.~T.} \bibnamefont{{Edwards}}},
  \bibinfo{author}{\bibfnamefont{G.~B.} \bibnamefont{{Hobbs}}},
  \bibnamefont{and} \bibinfo{author}{\bibfnamefont{R.~N.}
  \bibnamefont{{Manchester}}}, \bibinfo{journal}{\mnras}
  \textbf{\bibinfo{volume}{372}}, \bibinfo{pages}{1549} (\bibinfo{year}{2006}),
  \eprint{astro-ph/0607664}.

\bibitem[{\citenamefont{{Alpar} et~al.}(1986)\citenamefont{{Alpar},
  {Nandkumar}, and {Pines}}}]{Alpar1986}
\bibinfo{author}{\bibfnamefont{M.~A.} \bibnamefont{{Alpar}}},
  \bibinfo{author}{\bibfnamefont{R.}~\bibnamefont{{Nandkumar}}},
  \bibnamefont{and} \bibinfo{author}{\bibfnamefont{D.}~\bibnamefont{{Pines}}},
  \bibinfo{journal}{Astrophys. J} \textbf{\bibinfo{volume}{311}},
  \bibinfo{pages}{197} (\bibinfo{year}{1986}).

\bibitem[{\citenamefont{{Cheng}}(1987)}]{Cheng1987}
\bibinfo{author}{\bibfnamefont{K.~S.} \bibnamefont{{Cheng}}},
  \bibinfo{journal}{Astrophys. J} \textbf{\bibinfo{volume}{321}},
  \bibinfo{pages}{799} (\bibinfo{year}{1987}).

\bibitem[{\citenamefont{{Cordes}}(1993)}]{Cordes1993}
\bibinfo{author}{\bibfnamefont{J.~M.} \bibnamefont{{Cordes}}}, in
  \emph{\bibinfo{booktitle}{Planets Around Pulsars}}, edited by
  \bibinfo{editor}{\bibfnamefont{J.~A.} \bibnamefont{{Phillips}}},
  \bibinfo{editor}{\bibfnamefont{S.~E.} \bibnamefont{{Thorsett}}},
  \bibnamefont{and} \bibinfo{editor}{\bibfnamefont{S.~R.}
  \bibnamefont{{Kulkarni}}} (\bibinfo{year}{1993}), vol.~\bibinfo{volume}{36}
  of \emph{\bibinfo{series}{Astronomical Society of the Pacific Conference
  Series}}, pp. \bibinfo{pages}{43--60}.

\bibitem[{\citenamefont{{Cordes} and {Shannon}}(2008)}]{Cordes2008}
\bibinfo{author}{\bibfnamefont{J.~M.} \bibnamefont{{Cordes}}} \bibnamefont{and}
  \bibinfo{author}{\bibfnamefont{R.~M.} \bibnamefont{{Shannon}}},
  \bibinfo{journal}{Astrophys. J} \textbf{\bibinfo{volume}{682}},
  \bibinfo{eid}{1152-1165} (\bibinfo{year}{2008}), \eprint{astro-ph/0605145}.

\bibitem[{\citenamefont{{Shannon} et~al.}(2013)\citenamefont{{Shannon},
  {Cordes}, {Metcalfe}, {Lazio}, {Cognard}, {Desvignes}, {Janssen}, {Jessner},
  {Kramer}, {Lazaridis} et~al.}}]{Shannon2013}
\bibinfo{author}{\bibfnamefont{R.~M.} \bibnamefont{{Shannon}}},
  \bibinfo{author}{\bibfnamefont{J.~M.} \bibnamefont{{Cordes}}},
  \bibinfo{author}{\bibfnamefont{T.~S.} \bibnamefont{{Metcalfe}}},
  \bibinfo{author}{\bibfnamefont{T.~J.~W.} \bibnamefont{{Lazio}}},
  \bibinfo{author}{\bibfnamefont{I.}~\bibnamefont{{Cognard}}},
  \bibinfo{author}{\bibfnamefont{G.}~\bibnamefont{{Desvignes}}},
  \bibinfo{author}{\bibfnamefont{G.~H.} \bibnamefont{{Janssen}}},
  \bibinfo{author}{\bibfnamefont{A.}~\bibnamefont{{Jessner}}},
  \bibinfo{author}{\bibfnamefont{M.}~\bibnamefont{{Kramer}}},
  \bibinfo{author}{\bibfnamefont{K.}~\bibnamefont{{Lazaridis}}},
  \bibnamefont{et~al.}, \bibinfo{journal}{Astrophys. J}
  \textbf{\bibinfo{volume}{766}}, \bibinfo{eid}{5} (\bibinfo{year}{2013}),
  \eprint{1301.6429}.

\bibitem[{\citenamefont{{Shannon} et~al.}(2014)\citenamefont{{Shannon},
  {Os{\l}owski}, {Dai}, {Bailes}, {Hobbs}, {Manchester}, {van Straten},
  {Raithel}, {Ravi}, {Toomey} et~al.}}]{Shannon2014}
\bibinfo{author}{\bibfnamefont{R.~M.} \bibnamefont{{Shannon}}},
  \bibinfo{author}{\bibfnamefont{S.}~\bibnamefont{{Os{\l}owski}}},
  \bibinfo{author}{\bibfnamefont{S.}~\bibnamefont{{Dai}}},
  \bibinfo{author}{\bibfnamefont{M.}~\bibnamefont{{Bailes}}},
  \bibinfo{author}{\bibfnamefont{G.}~\bibnamefont{{Hobbs}}},
  \bibinfo{author}{\bibfnamefont{R.~N.} \bibnamefont{{Manchester}}},
  \bibinfo{author}{\bibfnamefont{W.}~\bibnamefont{{van Straten}}},
  \bibinfo{author}{\bibfnamefont{C.~A.} \bibnamefont{{Raithel}}},
  \bibinfo{author}{\bibfnamefont{V.}~\bibnamefont{{Ravi}}},
  \bibinfo{author}{\bibfnamefont{L.}~\bibnamefont{{Toomey}}},
  \bibnamefont{et~al.}, \bibinfo{journal}{\mnras}
  \textbf{\bibinfo{volume}{443}}, \bibinfo{pages}{1463} (\bibinfo{year}{2014}),
  \eprint{1406.4716}.

\bibitem[{\citenamefont{{Coles} et~al.}(2011)\citenamefont{{Coles}, {Hobbs},
  {Champion}, {Manchester}, and {Verbiest}}}]{Coles2011}
\bibinfo{author}{\bibfnamefont{W.}~\bibnamefont{{Coles}}},
  \bibinfo{author}{\bibfnamefont{G.}~\bibnamefont{{Hobbs}}},
  \bibinfo{author}{\bibfnamefont{D.~J.} \bibnamefont{{Champion}}},
  \bibinfo{author}{\bibfnamefont{R.~N.} \bibnamefont{{Manchester}}},
  \bibnamefont{and} \bibinfo{author}{\bibfnamefont{J.~P.~W.}
  \bibnamefont{{Verbiest}}}, \bibinfo{journal}{\mnras}
  \textbf{\bibinfo{volume}{418}}, \bibinfo{pages}{561} (\bibinfo{year}{2011}),
  \eprint{1107.5366}.

\bibitem[{\citenamefont{{Boynton} et~al.}(1972)\citenamefont{{Boynton},
  {Groth}, {Hutchinson}, {Nanos}, {Partridge}, and {Wilkinson}}}]{Boynton1972}
\bibinfo{author}{\bibfnamefont{P.~E.} \bibnamefont{{Boynton}}},
  \bibinfo{author}{\bibfnamefont{E.~J.} \bibnamefont{{Groth}}},
  \bibinfo{author}{\bibfnamefont{D.~P.} \bibnamefont{{Hutchinson}}},
  \bibinfo{author}{\bibfnamefont{G.~P.} \bibnamefont{{Nanos}},
  \bibfnamefont{Jr.}}, \bibinfo{author}{\bibfnamefont{R.~B.}
  \bibnamefont{{Partridge}}}, \bibnamefont{and}
  \bibinfo{author}{\bibfnamefont{D.~T.} \bibnamefont{{Wilkinson}}},
  \bibinfo{journal}{Astrophys. J} \textbf{\bibinfo{volume}{175}},
  \bibinfo{pages}{217} (\bibinfo{year}{1972}).

\bibitem[{\citenamefont{{Groth}}(1975)}]{Groth1975ApJS}
\bibinfo{author}{\bibfnamefont{E.~J.} \bibnamefont{{Groth}}},
  \bibinfo{journal}{\apjs} \textbf{\bibinfo{volume}{29}}
  (\bibinfo{year}{1975}).

\bibitem[{\citenamefont{{Cordes} and {Helfand}}(1980)}]{Cordes1980}
\bibinfo{author}{\bibfnamefont{J.~M.} \bibnamefont{{Cordes}}} \bibnamefont{and}
  \bibinfo{author}{\bibfnamefont{D.~J.} \bibnamefont{{Helfand}}},
  \bibinfo{journal}{Astrophys. J} \textbf{\bibinfo{volume}{239}},
  \bibinfo{pages}{640} (\bibinfo{year}{1980}).

\bibitem[{\citenamefont{{Cordes} and {Downs}}(1985)}]{Cordes1985}
\bibinfo{author}{\bibfnamefont{J.~M.} \bibnamefont{{Cordes}}} \bibnamefont{and}
  \bibinfo{author}{\bibfnamefont{G.~S.} \bibnamefont{{Downs}}},
  \bibinfo{journal}{\apjs} \textbf{\bibinfo{volume}{59}}, \bibinfo{pages}{343}
  (\bibinfo{year}{1985}).

\bibitem[{\citenamefont{{D'Alessandro}
  et~al.}(1995)\citenamefont{{D'Alessandro}, {McCulloch}, {Hamilton}, and
  {Deshpande}}}]{D'Alessandro1995}
\bibinfo{author}{\bibfnamefont{F.}~\bibnamefont{{D'Alessandro}}},
  \bibinfo{author}{\bibfnamefont{P.~M.} \bibnamefont{{McCulloch}}},
  \bibinfo{author}{\bibfnamefont{P.~A.} \bibnamefont{{Hamilton}}},
  \bibnamefont{and} \bibinfo{author}{\bibfnamefont{A.~A.}
  \bibnamefont{{Deshpande}}}, \bibinfo{journal}{\mnras}
  \textbf{\bibinfo{volume}{277}}, \bibinfo{pages}{1033} (\bibinfo{year}{1995}).

\bibitem[{\citenamefont{{Lyne} et~al.}(2010)\citenamefont{{Lyne}, {Hobbs},
  {Kramer}, {Stairs}, and {Stappers}}}]{Lyne2010}
\bibinfo{author}{\bibfnamefont{A.}~\bibnamefont{{Lyne}}},
  \bibinfo{author}{\bibfnamefont{G.}~\bibnamefont{{Hobbs}}},
  \bibinfo{author}{\bibfnamefont{M.}~\bibnamefont{{Kramer}}},
  \bibinfo{author}{\bibfnamefont{I.}~\bibnamefont{{Stairs}}}, \bibnamefont{and}
  \bibinfo{author}{\bibfnamefont{B.}~\bibnamefont{{Stappers}}},
  \bibinfo{journal}{Science} \textbf{\bibinfo{volume}{329}},
  \bibinfo{pages}{408} (\bibinfo{year}{2010}), \eprint{1006.5184}.

\bibitem[{\citenamefont{{Hobbs} et~al.}(2010)\citenamefont{{Hobbs}, {Lyne}, and
  {Kramer}}}]{Hobbs2010}
\bibinfo{author}{\bibfnamefont{G.}~\bibnamefont{{Hobbs}}},
  \bibinfo{author}{\bibfnamefont{A.~G.} \bibnamefont{{Lyne}}},
  \bibnamefont{and} \bibinfo{author}{\bibfnamefont{M.}~\bibnamefont{{Kramer}}},
  \bibinfo{journal}{\mnras} \textbf{\bibinfo{volume}{402}},
  \bibinfo{pages}{1027} (\bibinfo{year}{2010}), \eprint{0912.4537}.

\bibitem[{\citenamefont{{Shabanova} et~al.}(2013)\citenamefont{{Shabanova},
  {Pugachev}, and {Lapaev}}}]{Shabanova2013}
\bibinfo{author}{\bibfnamefont{T.~V.} \bibnamefont{{Shabanova}}},
  \bibinfo{author}{\bibfnamefont{V.~D.} \bibnamefont{{Pugachev}}},
  \bibnamefont{and} \bibinfo{author}{\bibfnamefont{K.~A.}
  \bibnamefont{{Lapaev}}}, \bibinfo{journal}{Astrophys. J}
  \textbf{\bibinfo{volume}{775}}, \bibinfo{eid}{2} (\bibinfo{year}{2013}),
  \eprint{1307.0297}.

\bibitem[{\citenamefont{{Pines} and {Shaham}}(1972)}]{Pines1972}
\bibinfo{author}{\bibfnamefont{D.}~\bibnamefont{{Pines}}} \bibnamefont{and}
  \bibinfo{author}{\bibfnamefont{J.}~\bibnamefont{{Shaham}}},
  \bibinfo{journal}{Nature Physical Science} \textbf{\bibinfo{volume}{235}},
  \bibinfo{pages}{43} (\bibinfo{year}{1972}).

\bibitem[{\citenamefont{{Packard}}(1972)}]{Packard1972}
\bibinfo{author}{\bibfnamefont{R.~E.} \bibnamefont{{Packard}}},
  \bibinfo{journal}{Physical Review Letters} \textbf{\bibinfo{volume}{28}},
  \bibinfo{pages}{1080} (\bibinfo{year}{1972}).

\bibitem[{\citenamefont{{Lamb} et~al.}(1978)\citenamefont{{Lamb}, {Pines}, and
  {Shaham}}}]{Lamb1978}
\bibinfo{author}{\bibfnamefont{F.~K.} \bibnamefont{{Lamb}}},
  \bibinfo{author}{\bibfnamefont{D.}~\bibnamefont{{Pines}}}, \bibnamefont{and}
  \bibinfo{author}{\bibfnamefont{J.}~\bibnamefont{{Shaham}}},
  \bibinfo{journal}{Astrophys. J} \textbf{\bibinfo{volume}{224}},
  \bibinfo{pages}{969} (\bibinfo{year}{1978}).

\bibitem[{\citenamefont{{Cordes}}(2013)}]{Cordes2013}
\bibinfo{author}{\bibfnamefont{J.~M.} \bibnamefont{{Cordes}}},
  \bibinfo{journal}{Astrophys. J} \textbf{\bibinfo{volume}{775}},
  \bibinfo{eid}{47} (\bibinfo{year}{2013}), \eprint{1304.5803}.

\bibitem[{\citenamefont{{D'Allesandro} and
  {McCulloch}}(1997)}]{D'Allesandro1997}
\bibinfo{author}{\bibfnamefont{F.}~\bibnamefont{{D'Allesandro}}}
  \bibnamefont{and} \bibinfo{author}{\bibfnamefont{P.~M.}
  \bibnamefont{{McCulloch}}}, \bibinfo{journal}{\mnras}
  \textbf{\bibinfo{volume}{292}}, \bibinfo{pages}{879} (\bibinfo{year}{1997}).

\bibitem[{\citenamefont{{van Horn}}(1980)}]{Horn1980}
\bibinfo{author}{\bibfnamefont{H.~M.} \bibnamefont{{van Horn}}},
  \bibinfo{journal}{Astrophys. J} \textbf{\bibinfo{volume}{236}},
  \bibinfo{pages}{899} (\bibinfo{year}{1980}).

\bibitem[{\citenamefont{{Shaham}}(1977)}]{Shaham1977}
\bibinfo{author}{\bibfnamefont{J.}~\bibnamefont{{Shaham}}},
  \bibinfo{journal}{Astrophys. J} \textbf{\bibinfo{volume}{214}},
  \bibinfo{pages}{251} (\bibinfo{year}{1977}).

\bibitem[{\citenamefont{{Stairs} et~al.}(2000)\citenamefont{{Stairs}, {Lyne},
  and {Shemar}}}]{Stairs2000}
\bibinfo{author}{\bibfnamefont{I.~H.} \bibnamefont{{Stairs}}},
  \bibinfo{author}{\bibfnamefont{A.~G.} \bibnamefont{{Lyne}}},
  \bibnamefont{and} \bibinfo{author}{\bibfnamefont{S.~L.}
  \bibnamefont{{Shemar}}}, \bibinfo{journal}{\nat}
  \textbf{\bibinfo{volume}{406}}, \bibinfo{pages}{484} (\bibinfo{year}{2000}).

\bibitem[{\citenamefont{{Jones}}(2012)}]{Jones2012}
\bibinfo{author}{\bibfnamefont{D.~I.} \bibnamefont{{Jones}}},
  \bibinfo{journal}{\mnras} \textbf{\bibinfo{volume}{420}},
  \bibinfo{pages}{2325} (\bibinfo{year}{2012}), \eprint{1107.3503}.

\bibitem[{\citenamefont{{Wolszczan}}(1994)}]{Wolszczan1994}
\bibinfo{author}{\bibfnamefont{A.}~\bibnamefont{{Wolszczan}}},
  \bibinfo{journal}{Science} \textbf{\bibinfo{volume}{264}},
  \bibinfo{pages}{538} (\bibinfo{year}{1994}).

\bibitem[{\citenamefont{{Kramer} et~al.}(2006)\citenamefont{{Kramer}, {Lyne},
  {O'Brien}, {Jordan}, and {Lorimer}}}]{Kramer2006}
\bibinfo{author}{\bibfnamefont{M.}~\bibnamefont{{Kramer}}},
  \bibinfo{author}{\bibfnamefont{A.~G.} \bibnamefont{{Lyne}}},
  \bibinfo{author}{\bibfnamefont{J.~T.} \bibnamefont{{O'Brien}}},
  \bibinfo{author}{\bibfnamefont{C.~A.} \bibnamefont{{Jordan}}},
  \bibnamefont{and} \bibinfo{author}{\bibfnamefont{D.~R.}
  \bibnamefont{{Lorimer}}}, \bibinfo{journal}{Science}
  \textbf{\bibinfo{volume}{312}}, \bibinfo{pages}{549} (\bibinfo{year}{2006}),
  \eprint{astro-ph/0604605}.

\bibitem[{\citenamefont{{Link}}(2003)}]{Link2003}
\bibinfo{author}{\bibfnamefont{B.}~\bibnamefont{{Link}}},
  \bibinfo{journal}{Physical Review Letters} \textbf{\bibinfo{volume}{91}},
  \bibinfo{eid}{101101} (\bibinfo{year}{2003}), \eprint{astro-ph/0302441}.

\bibitem[{\citenamefont{{Zhang} and {Xie}}(2012)}]{Zhang2012}
\bibinfo{author}{\bibfnamefont{S.-N.} \bibnamefont{{Zhang}}} \bibnamefont{and}
  \bibinfo{author}{\bibfnamefont{Y.}~\bibnamefont{{Xie}}},
  \bibinfo{journal}{Astrophys. J} \textbf{\bibinfo{volume}{761}},
  \bibinfo{eid}{102} (\bibinfo{year}{2012}), \eprint{1209.2478}.

\bibitem[{\citenamefont{{Goldreich} and {Reisenegger}}(1992)}]{Goldreich92}
\bibinfo{author}{\bibfnamefont{P.}~\bibnamefont{{Goldreich}}} \bibnamefont{and}
  \bibinfo{author}{\bibfnamefont{A.}~\bibnamefont{{Reisenegger}}},
  \bibinfo{journal}{Astrophys. J.} \textbf{\bibinfo{volume}{395}},
  \bibinfo{pages}{250} (\bibinfo{year}{1992}).

\bibitem[{\citenamefont{{Pons} and {Rea}}(2012)}]{Pons2012}
\bibinfo{author}{\bibfnamefont{J.~A.} \bibnamefont{{Pons}}} \bibnamefont{and}
  \bibinfo{author}{\bibfnamefont{N.}~\bibnamefont{{Rea}}},
  \bibinfo{journal}{Astrophys. J.} \textbf{\bibinfo{volume}{750}},
  \bibinfo{eid}{L6} (\bibinfo{year}{2012}), \eprint{1203.4506}.

\bibitem[{\citenamefont{{Shoenberg}}(1984)}]{Shoenberg84}
\bibinfo{author}{\bibfnamefont{D.}~\bibnamefont{{Shoenberg}}},
  \emph{\bibinfo{title}{{Magnetic oscillation in metals}}}
  (\bibinfo{year}{1984}), pp. \bibinfo{pages}{32--66}.

\bibitem[{\citenamefont{{Condon}}(1966)}]{Condon1966}
\bibinfo{author}{\bibfnamefont{J.~H.} \bibnamefont{{Condon}}},
  \bibinfo{journal}{Physical Review} \textbf{\bibinfo{volume}{145}},
  \bibinfo{pages}{526} (\bibinfo{year}{1966}).

\bibitem[{\citenamefont{{Gordon} et~al.}(2003)\citenamefont{{Gordon}, {Vagner},
  and {Wyder}}}]{Gordon2003}
\bibinfo{author}{\bibfnamefont{A.}~\bibnamefont{{Gordon}}},
  \bibinfo{author}{\bibfnamefont{I.~D.} \bibnamefont{{Vagner}}},
  \bibnamefont{and} \bibinfo{author}{\bibfnamefont{P.}~\bibnamefont{{Wyder}}},
  \bibinfo{journal}{Advances in Physics} \textbf{\bibinfo{volume}{52}},
  \bibinfo{pages}{385} (\bibinfo{year}{2003}).

\bibitem[{\citenamefont{{Blandford} and {Hernquist}}(1982)}]{Blandford82}
\bibinfo{author}{\bibfnamefont{R.~D.} \bibnamefont{{Blandford}}}
  \bibnamefont{and}
  \bibinfo{author}{\bibfnamefont{L.}~\bibnamefont{{Hernquist}}},
  \bibinfo{journal}{Journal of Physics C Solid State Physics}
  \textbf{\bibinfo{volume}{15}}, \bibinfo{pages}{6233} (\bibinfo{year}{1982}).

\bibitem[{\citenamefont{{Suh} and {Mathews}}(2010)}]{Suh10}
\bibinfo{author}{\bibfnamefont{I.-S.} \bibnamefont{{Suh}}} \bibnamefont{and}
  \bibinfo{author}{\bibfnamefont{G.~J.} \bibnamefont{{Mathews}}},
  \bibinfo{journal}{Astrophys. J.} \textbf{\bibinfo{volume}{717}},
  \bibinfo{pages}{843} (\bibinfo{year}{2010}), \eprint{1005.2139}.

\bibitem[{\citenamefont{{Wang} et~al.}(2013)\citenamefont{{Wang}, {L{\"u}},
  {Zhu}, and {Huo}}}]{Wang2013}
\bibinfo{author}{\bibfnamefont{Z.}~\bibnamefont{{Wang}}},
  \bibinfo{author}{\bibfnamefont{G.}~\bibnamefont{{L{\"u}}}},
  \bibinfo{author}{\bibfnamefont{C.}~\bibnamefont{{Zhu}}}, \bibnamefont{and}
  \bibinfo{author}{\bibfnamefont{W.}~\bibnamefont{{Huo}}},
  \bibinfo{journal}{Astrophys. J} \textbf{\bibinfo{volume}{773}},
  \bibinfo{eid}{160} (\bibinfo{year}{2013}), \eprint{1303.2442}.

\bibitem[{\citenamefont{{Lyne}}(2013)}]{Lune2013}
\bibinfo{author}{\bibfnamefont{A.}~\bibnamefont{{Lyne}}}, in
  \emph{\bibinfo{booktitle}{Neutron Stars and Pulsars: Challenges and
  Opportunities after 80 years}}, edited by
  \bibinfo{editor}{\bibfnamefont{J.}~\bibnamefont{{van Leeuwen}}}
  (\bibinfo{year}{2013}), vol. \bibinfo{volume}{291} of
  \emph{\bibinfo{series}{IAU Symposium}}, pp. \bibinfo{pages}{183--188},
  \eprint{1212.2250}.

\bibitem[{\citenamefont{{Johnson} and {Lippmann}}(1949)}]{Jonhson49}
\bibinfo{author}{\bibfnamefont{M.~H.} \bibnamefont{{Johnson}}}
  \bibnamefont{and} \bibinfo{author}{\bibfnamefont{B.~A.}
  \bibnamefont{{Lippmann}}}, \bibinfo{journal}{Physical Review}
  \textbf{\bibinfo{volume}{76}}, \bibinfo{pages}{828} (\bibinfo{year}{1949}).

\bibitem[{\citenamefont{{Dingle}}(1952)}]{dingle1952}
\bibinfo{author}{\bibfnamefont{R.~B.} \bibnamefont{{Dingle}}},
  \bibinfo{journal}{Royal Society of London Proceedings Series A}
  \textbf{\bibinfo{volume}{211}}, \bibinfo{pages}{500} (\bibinfo{year}{1952}).

\bibitem[{\citenamefont{{Chamel} et~al.}(2012)\citenamefont{{Chamel}, {Pavlov},
  {Mihailov}, {Velchev}, {Stoyanov}, {Mutafchieva}, {Ivanovich}, {Pearson}, and
  {Goriely}}}]{Chamel2012}
\bibinfo{author}{\bibfnamefont{N.}~\bibnamefont{{Chamel}}},
  \bibinfo{author}{\bibfnamefont{R.~L.} \bibnamefont{{Pavlov}}},
  \bibinfo{author}{\bibfnamefont{L.~M.} \bibnamefont{{Mihailov}}},
  \bibinfo{author}{\bibfnamefont{C.~J.} \bibnamefont{{Velchev}}},
  \bibinfo{author}{\bibfnamefont{Z.~K.} \bibnamefont{{Stoyanov}}},
  \bibinfo{author}{\bibfnamefont{Y.~D.} \bibnamefont{{Mutafchieva}}},
  \bibinfo{author}{\bibfnamefont{M.~D.} \bibnamefont{{Ivanovich}}},
  \bibinfo{author}{\bibfnamefont{J.~M.} \bibnamefont{{Pearson}}},
  \bibnamefont{and}
  \bibinfo{author}{\bibfnamefont{S.}~\bibnamefont{{Goriely}}},
  \bibinfo{journal}{\prc} \textbf{\bibinfo{volume}{86}}, \bibinfo{eid}{055804}
  (\bibinfo{year}{2012}), \eprint{1210.5874}.

\bibitem[{\citenamefont{{Hernquist}}(1985)}]{Hernquist1985}
\bibinfo{author}{\bibfnamefont{L.}~\bibnamefont{{Hernquist}}},
  \bibinfo{journal}{\mnras} \textbf{\bibinfo{volume}{213}},
  \bibinfo{pages}{313} (\bibinfo{year}{1985}).

\bibitem[{\citenamefont{{Duncan} and {Thompson}}(1992)}]{Duncan92}
\bibinfo{author}{\bibfnamefont{R.~C.} \bibnamefont{{Duncan}}} \bibnamefont{and}
  \bibinfo{author}{\bibfnamefont{C.}~\bibnamefont{{Thompson}}},
  \bibinfo{journal}{Astrophys. J} \textbf{\bibinfo{volume}{392}},
  \bibinfo{pages}{L9} (\bibinfo{year}{1992}).

\bibitem[{\citenamefont{{Cumming} et~al.}(2004)\citenamefont{{Cumming},
  {Arras}, and {Zweibel}}}]{cumming2004}
\bibinfo{author}{\bibfnamefont{A.}~\bibnamefont{{Cumming}}},
  \bibinfo{author}{\bibfnamefont{P.}~\bibnamefont{{Arras}}}, \bibnamefont{and}
  \bibinfo{author}{\bibfnamefont{E.}~\bibnamefont{{Zweibel}}},
  \bibinfo{journal}{Astrophys. J.} \textbf{\bibinfo{volume}{609}},
  \bibinfo{pages}{999} (\bibinfo{year}{2004}), \eprint{arXiv:astro-ph/0402392}.

\bibitem[{\citenamefont{{Itoh} and {Kohyama}}(1993)}]{Itoh1993}
\bibinfo{author}{\bibfnamefont{N.}~\bibnamefont{{Itoh}}} \bibnamefont{and}
  \bibinfo{author}{\bibfnamefont{Y.}~\bibnamefont{{Kohyama}}},
  \bibinfo{journal}{Astrophys. J} \textbf{\bibinfo{volume}{404}},
  \bibinfo{pages}{268} (\bibinfo{year}{1993}).

\bibitem[{\citenamefont{{Jones}}(2001)}]{Jones2001}
\bibinfo{author}{\bibfnamefont{P.~B.} \bibnamefont{{Jones}}},
  \bibinfo{journal}{MNRAS} \textbf{\bibinfo{volume}{321}}, \bibinfo{pages}{167}
  (\bibinfo{year}{2001}).

\bibitem[{\citenamefont{{Condon} and {Walstedt}}(1968)}]{Condon1968}
\bibinfo{author}{\bibfnamefont{J.~H.} \bibnamefont{{Condon}}} \bibnamefont{and}
  \bibinfo{author}{\bibfnamefont{R.~E.} \bibnamefont{{Walstedt}}},
  \bibinfo{journal}{Physical Review Letters} \textbf{\bibinfo{volume}{21}},
  \bibinfo{pages}{612} (\bibinfo{year}{1968}).

\bibitem[{\citenamefont{{Wang} et~al.}(2016)\citenamefont{{Wang}, {L{\"u}},
  {Zhu}, and {Wu}}}]{Wang2016}
\bibinfo{author}{\bibfnamefont{Z.}~\bibnamefont{{Wang}}},
  \bibinfo{author}{\bibfnamefont{G.}~\bibnamefont{{L{\"u}}}},
  \bibinfo{author}{\bibfnamefont{C.}~\bibnamefont{{Zhu}}}, \bibnamefont{and}
  \bibinfo{author}{\bibfnamefont{B.}~\bibnamefont{{Wu}}},
  \bibinfo{journal}{\pasp} \textbf{\bibinfo{volume}{128}},
  \bibinfo{pages}{104201} (\bibinfo{year}{2016}).

\bibitem[{\citenamefont{{Wong}}(2001)}]{Wong2001}
\bibinfo{author}{\bibfnamefont{R.}~\bibnamefont{{Wong}}},
  \emph{\bibinfo{title}{{Asymptotic Approximations of Integrals}}}
  (\bibinfo{year}{2001}), pp. \bibinfo{pages}{423--475}.

\bibitem[{\citenamefont{{Egorov}}(2010)}]{Egorov2010}
\bibinfo{author}{\bibfnamefont{V.~S.} \bibnamefont{{Egorov}}},
  \bibinfo{journal}{Physics Uspekhi} \textbf{\bibinfo{volume}{53}},
  \bibinfo{pages}{755} (\bibinfo{year}{2010}).

\bibitem[{\citenamefont{{Horowitz} and {Kadau}}(2009)}]{Horowitz2009}
\bibinfo{author}{\bibfnamefont{C.~J.} \bibnamefont{{Horowitz}}}
  \bibnamefont{and} \bibinfo{author}{\bibfnamefont{K.}~\bibnamefont{{Kadau}}},
  \bibinfo{journal}{Physical Review Letters} \textbf{\bibinfo{volume}{102}},
  \bibinfo{eid}{191102} (\bibinfo{year}{2009}), \eprint{0904.1986}.

\bibitem[{\citenamefont{{Narayan}}(1996)}]{Narayan1996}
\bibinfo{author}{\bibfnamefont{O.}~\bibnamefont{{Narayan}}},
  \bibinfo{journal}{Physical Review Letters} \textbf{\bibinfo{volume}{77}},
  \bibinfo{pages}{3855} (\bibinfo{year}{1996}),
  \eprint{arXiv:cond-mat/9610053}.

\bibitem[{\citenamefont{{Rheinhardt} and {Geppert}}(2002)}]{Rheinhardt2002}
\bibinfo{author}{\bibfnamefont{M.}~\bibnamefont{{Rheinhardt}}}
  \bibnamefont{and}
  \bibinfo{author}{\bibfnamefont{U.}~\bibnamefont{{Geppert}}},
  \bibinfo{journal}{Physical Review Letters} \textbf{\bibinfo{volume}{88}},
  \bibinfo{eid}{101103} (\bibinfo{year}{2002}).

\bibitem[{\citenamefont{{Pons} and {Geppert}}(2010)}]{Pons2010}
\bibinfo{author}{\bibfnamefont{J.~A.} \bibnamefont{{Pons}}} \bibnamefont{and}
  \bibinfo{author}{\bibfnamefont{U.}~\bibnamefont{{Geppert}}},
  \bibinfo{journal}{\aap} \textbf{\bibinfo{volume}{513}}, \bibinfo{eid}{L12}
  (\bibinfo{year}{2010}), \eprint{1004.1054}.

\bibitem[{\citenamefont{{Cizeau} et~al.}(1997)\citenamefont{{Cizeau},
  {Zapperi}, {Durin}, and {Stanley}}}]{Cizeau1997}
\bibinfo{author}{\bibfnamefont{P.}~\bibnamefont{{Cizeau}}},
  \bibinfo{author}{\bibfnamefont{S.}~\bibnamefont{{Zapperi}}},
  \bibinfo{author}{\bibfnamefont{G.}~\bibnamefont{{Durin}}}, \bibnamefont{and}
  \bibinfo{author}{\bibfnamefont{H.~E.} \bibnamefont{{Stanley}}},
  \bibinfo{journal}{Physical Review Letters} \textbf{\bibinfo{volume}{79}},
  \bibinfo{pages}{4669} (\bibinfo{year}{1997}), \eprint{cond-mat/9709300}.

\bibitem[{\citenamefont{{Blaes} et~al.}(1989)\citenamefont{{Blaes},
  {Blandford}, {Goldreich}, and {Madau}}}]{Blaes1989}
\bibinfo{author}{\bibfnamefont{O.}~\bibnamefont{{Blaes}}},
  \bibinfo{author}{\bibfnamefont{R.}~\bibnamefont{{Blandford}}},
  \bibinfo{author}{\bibfnamefont{P.}~\bibnamefont{{Goldreich}}},
  \bibnamefont{and} \bibinfo{author}{\bibfnamefont{P.}~\bibnamefont{{Madau}}},
  \bibinfo{journal}{Astrophys. J} \textbf{\bibinfo{volume}{343}},
  \bibinfo{pages}{839} (\bibinfo{year}{1989}).

\end{thebibliography}

\end{document}